\documentclass[aps,10pt,prb,twocolumn, superscriptaddress]{revtex4-2}
\usepackage{graphicx}
\usepackage{amssymb, amsmath}
\usepackage{amsfonts}
\usepackage{subfigure}
\usepackage{epstopdf}
\usepackage{pbox}
\usepackage{xcolor}

\usepackage{setspace}

\begin{document}

\title{Magnetic structure and spin dynamics of the quasi-2D antiferromagnet Zn-doped copper pyrovanadate}

\author{G. Gitgeatpong}
\email[]{ganatee.g@pnru.ac.th}
\affiliation{Faculty of Science and Technology, Phranakhon Rajabhat University, Bangkok 10220, Thailand}
\affiliation{Thailand Center of Excellence in Physics, Ministry of Higher Education, Science, Research and Innovation, 328 Si Ayutthaya Road, Bangkok 10400, Thailand}

\author{Y. Zhao}
\affiliation{Department of Materials Science and Engineering, University of Maryland, College Park, Maryland 20742, USA.}
\affiliation{NIST Center for Neutron Research, National Institute of Standards and Technology, Gaithersburg, Maryland 20899, USA.}

\author{J. A. Fernandez-Baca}
\affiliation{Neutron Scattering Division, Oak Ridge National Laboratory, Oak Ridge, Tennessee 37831, USA.}

\author{T.~Hong}
\affiliation{Neutron Scattering Division, Oak Ridge National Laboratory, Oak Ridge, Tennessee 37831, USA.}

\author{T.~J.~Sato}
\affiliation{IMRAM, Tohoku University, Sendai, Miyagi 980-8577, Japan.}

\author{P. Piyawongwatthana}
\affiliation{IMRAM, Tohoku University, Sendai, Miyagi 980-8577, Japan.}

\author{K. Nawa}
\affiliation{IMRAM, Tohoku University, Sendai, Miyagi 980-8577, Japan.}

\author{P. Saeaun}
\affiliation{Department of Physics, Faculty of Science, Mahidol University, Bangkok 10400, Thailand}

\author{K. Matan}
\email[]{kittiwit.mat@mahidol.ac.th}
\affiliation{Thailand Center of Excellence in Physics, Ministry of Higher Education, Science, Research and Innovation, 328 Si Ayutthaya Road, Bangkok 10400, Thailand}
\affiliation{Department of Physics, Faculty of Science, Mahidol University, Bangkok 10400, Thailand}

\date{\today}

\begin{abstract}
Magnetic properties of the antiferromagnet Zn$_{x}$Cu$_{2-x}$V$_2$O$_7$ (ZnCVO) with $x \approx$ 0.06 have been thoroughly investigated on powder and single-crystal samples. The crystal structure determination using powder x-ray and neutron diffraction confirms that our ZnCVO samples are isostructural with $\beta$-Cu$_{2}$V$_2$O$_7$ ($\beta$-CVO) with a small deviation in the lattice parameters. Macroscopic magnetic property measurements also confirm the similarity between the two compounds. The Cu$^{2+}$ spins were found to align along the crystallographic $c$-axis, antiparallel to their nearest neighbors connected by the leading exchange interaction $J_1$. Spin dynamics reveals a typical symmetric spin-wave dispersion with strong interactions in the $bc$-plane and weak interplane coupling. The exchange interaction analysis indicates that the spin network of ZnCVO is topologically consistent with the previous DFT prediction but the values of leading exchange interactions are contradictory. Furthermore, rather than the predicted 2D honeycomb structure, the spin network in ZnCVO could be better described by the anisotropic 2D spin network composing of $J_1$, $J_5$, and $J_6$ interactions, four bonds per one spin site, coupled by weak interplane interactions.

\end{abstract}

\maketitle

\section{Introduction}

The symmetry of solids plays an important role in determining crystal structure and the underlying physical properties, particularly a spin network, magnetic ground states, and spin dynamics in low-dimensional antiferromagnetic systems. According to Friedel's law~\cite{Sands} when the crystals have a center of symmetry at the origin, the structure factor for the $(hkl)$ and $(\bar{h} \bar{k} \bar{l})$ planes will result in the same intensity i.e., $|F(hkl)|^2$ = $|F(\bar{h} \bar{k} \bar{l})|^2$. The crystals that obey this rule are called {\sl centrosymmetric} crystals; otherwise, they are called {\sl non-centrosymmetric}. This rule also applies to the dispersion relation. In the non-centrosymmetric crystals, the system could present the uniform antisymmetric Dzyaloshinskii-Moriya (DM) interaction~\cite{Dzyaloshinsky, Moriya} between interacting magnetic spins, in which the asymmetric dispersion relation (nonreciprocal magnon) i.e., $E(k)~\neq~E(-k)$, is expected and experimentally observed~\cite{Hayami2016, Tokura2018, Sato2019,Gitgeatpong2017}. On the other hand, the asymmetric dispersion relation vanishes and is replaced by the conventional symmetric dispersion relation in the centrosymmetric crystals.

In our previous study~\cite{Gitgeatpong2017} on the non-centrosymmetric $\alpha$-Cu$_{2}$V$_2$O$_7$ ($\alpha$-CVO) or {\sl Blossite}, we surprisingly discovered the nonreciprocal magnon where the rare phenomenon of a bidirectional shift of the magnon dispersion was experimentally observed for the first time in an antiferromagnet. This discovery was a great proof of the theoretical prediction of the asymmetric dispersion relation in non-centrosymmetric crystals and raised our attention to Cu$_{2}$V$_2$O$_7$ system. There are three main polymorphs with a chemical formula Cu$_{2}$V$_2$O$_7$ i.e., $\alpha$, $\beta$, and $\gamma$. The $\gamma$-phase is more likely a complex high-temperature phase with the lowest crystal symmetry $P1$~\cite{Petrova2005}. A more related cousin phase to the $\alpha$-CVO is $\beta$-Cu$_{2}$V$_2$O$_7$ ($\beta$-CVO) or {\sl Ziesite} which is a centrosymmetric crystal. Both $\alpha$-CVO and $\beta$-CVO were naturally discovered at the summit crater of the Izalco volcano, El Salvador~\cite{Robinson1987, Hughes1980}. Despite the same chemical formula and the same nature of origin, the symmetry and magnetic properties of $\alpha$-CVO and $\beta$-CVO are quite different. We, therefore, extend our investigation from the non-centrosymmetric $\alpha$-CVO to the centrosymmetric $\beta$-CVO focusing on the magnetic properties and especially the spin dynamics.

The crystal structure of $\beta$-CVO is monoclinic with space group $C2/c$. The lattice parameters are $a$ = 7.685~\AA, $b$ = 8.007~\AA, $c$ = 10.09~\AA, and $\beta$ = 110.27$^\circ$~\cite{Mercurio1973, Hughes1980}. Unlike $\alpha$-CVO, the DM interaction is absent in $\beta$-CVO and thus the symmetric dispersion relation with $E(k) = E(-k)$ is expected. This system was first believed to be the antiferromagnetic 1D spin chain~\cite{Pommer2003, He2008a, Yashima2009} but the later DFT studies proposed the otherwise 2D honeycomb spin network~\cite{Tsirlin2010, Bhowal2017}. Here we performed a thorough experiment to investigate the magnetic structure as well as the spin-wave dispersion using state of the art neutron scattering technique to resolve this ambiguity. We chose Zn$_{x}$Cu$_{2-x}$V$_2$O$_7$ with $x \approx$ 0.06 (ZnCVO) as a prototypical sample because of its phase controllability. There were several reports on the synthesis of $\beta$-CVO samples, both powder~\cite{Krasnenko2008, Slobodin2009} and single-crystals~\cite{He2008a}. However, the $\alpha$ to $\beta$ phase transition temperatures were reported to be different~\cite{Clark1977, Slobodin2009} causing difficulty in growing the large-sized single-crystal for an inelastic neutron scattering study. Alternatively, Zn substitution on Cu sites can transform the formerly $\alpha$-Cu$_{2}$V$_2$O$_7$ to Zn$_{x}$Cu$_{2-x}$V$_2$O$_7$, which were reported to have the same crystal structure as $\beta$-CVO~\cite{Nord1985, Schindler1999, Pommer2003, Kataev2004, Salah2005, Sotojima2007}. There are many interesting aspects in the physical properties of these copper vanadate systems not only magnetic properties but also their negative thermal expansion~\cite{Wang2019, Sato2020, Shi2020} and photoelectrochemical properties~\cite{Muthamizh2020, Song2020}. Understanding the physics of $\beta$-CVO/ZnCVO could potentially lead to an insight into the low-dimensional quantum materials and their possible diverse applications.

The manuscript is organized as follows. We briefly start with the experimental details in Section~\ref{experiment} describing the sample preparations and the data collections. In Section~\ref{resultanddiscuss}, we allocate into four subsections. The first two subsections, \ref{sectionA} and \ref{sectionB}, will be discussing the crystal and magnetic structures, respectively, of ZnCVO and $\beta$-CVO samples. The next two subsections, \ref{sectionC} and \ref{sectionD}, will be devoted to the exchange interactions and spin network analysis. We finally end with the conclusion in Section~\ref{summary}.

\section{Experimental details}\label{experiment}
Powder samples of ZnCVO were prepared by the standard solid-state reaction from the stoichiometric ratio of ZnO, CuO, and V$_2$O$_5$. The mixture was ground and calcined repeatedly at the temperature between 600 - 650$^\circ$C in the air. Phase purity was checked by the powder x-ray diffraction. For comparison, a powder sample of the pure phase $\beta$-Cu$_2$V$_2$O$_7$ ($\beta$-CVO) was also prepared. The stoichiometric ratio of CuO, and V$_2$O$_5$ were mixed and ground thoroughly. The mixture was calcined and sintered at a temperature below 600$^\circ$C to avoid the $\alpha-\beta$ phase transition~\cite{Clark1977}, with intermediate grindings for a total of around 80 hours. The pure phase ZnCVO was used as a starting material for single-crystal growth using the vertical gradient furnace. The powder was put into a quartz tube and melted in the ambient air at around 850$^\circ$C before moving the molten sample down through the natural temperature gradient between 20$^\circ$C/cm~-~50$^\circ$C/cm with a rate of 1 cm/day. After the sample reaches the temperature of $\approx$ 600$^\circ$C, the crystals were then naturally cooled in the furnace to room temperature and mechanically extracted from quartz. 

The phase of the single crystals was first checked by powder x-ray diffraction on the ground crystals. Magnetic susceptibility measurements were done on a small piece of single-crystal by applying the magnetic field along two orthogonal directions i.e., $H \parallel a$ and $H \perp a$ using a superconducting quantum interference device (MPMS-XL, Quantum Design) with the field of 1 T. The obtained magnetic susceptibility data were analyzed and compared with the Quantum Monte Carlo simulation. Powder neutron diffraction data on both ZnCVO and $\beta$-CVO were collected at BT1, NIST Center for Neutron Research (NCNR), the USA for nuclear and magnetic structure determinations. Finally, inelastic neutron scattering experiments were done on a large piece of crystal ($m \approx$ 1.5 g). The crystal was aligned so that ($h,k,0$) was on the scattering plane. At the BT7 Double Focusing Thermal Triple Axis Spectrometer, NIST Center for Neutron Research (NCNR), USA, the scattered neutron energy was fixed 14.7 meV. The rocking scan was done on the major nuclear Bragg peak to qualify the crystallinity of the single crystal. The energy scans were collected at the base temperature along ($0,k,0$) and ($h,2,0$) directions over the broad range of the spin-wave dispersion. The energy scans around the magnetic zone center were done at the SPINS spectrometers, NCNR, and at the CTAX spectrometer, Oak Ridge National Laboratory, the USA with fixed scattered neutron energy of 5~meV to resolve the energy gap.

\section{Results and discussion}\label{resultanddiscuss}

\subsection{Crystal structure}\label{sectionA}

\begin{figure}
 \includegraphics[width=0.48\textwidth]{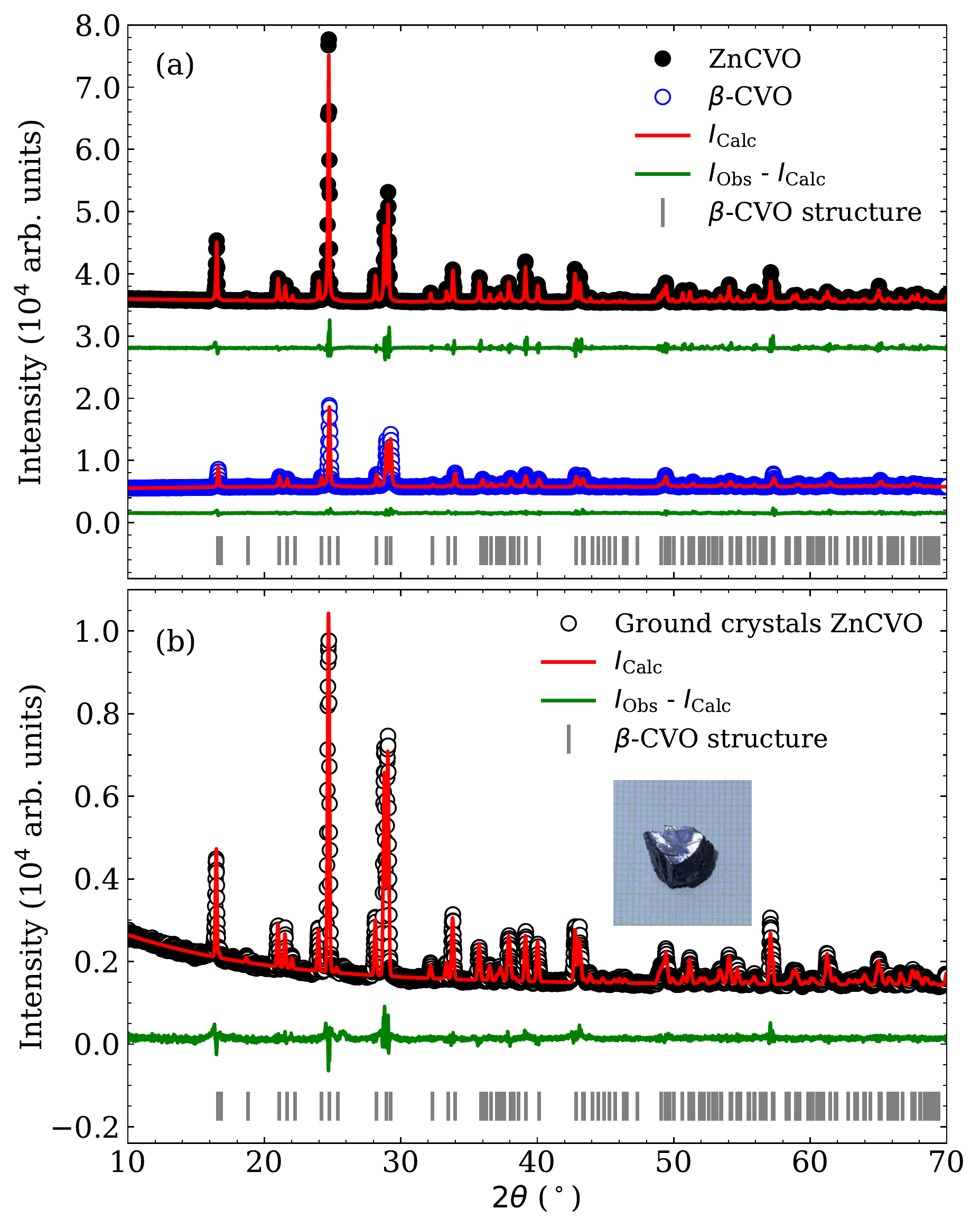}
 \caption{\label{fig1} (a) Powder x-ray diffraction patterns with the Rietveld refinements of the powder samples ZnCVO (black solid circle) and $\beta$-Cu$_2$V$_2$O$_7$ (blue open circle) collected at room temperature. (b) X-ray diffraction pattern with the Rietveld refinements of the ground single-crystals ZnCVO. In both panels, red lines are the calculated pattern, green lines are the difference between the observed and calculated patterns, and the vertical grey ticks represent the Bragg positions for $\beta$-Cu$_2$V$_2$O$_7$ structure. The inset shows a photograph of the obtained single crystal.}
 \end{figure}

\begin{figure}
 \includegraphics[width=0.48\textwidth]{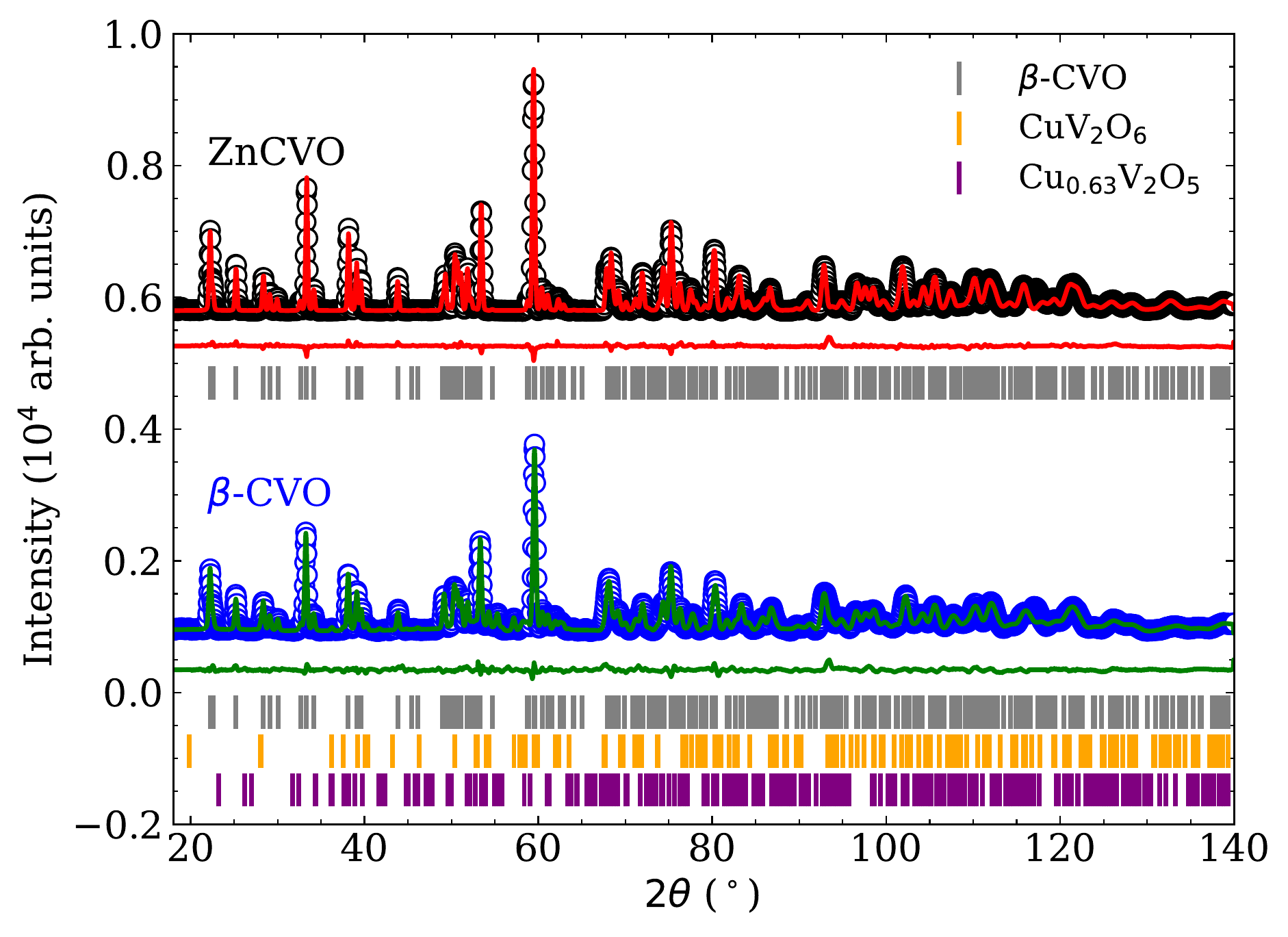}
 \caption{\label{fig2} Powder neutron diffraction patterns with the Rietveld refinements of the powder samples ZnCVO (black circle) and $\beta$-Cu$_2$V$_2$O$_7$ (blue circle) collected at $T$ = 30 K at the BT1 spectrometer, NCNR, USA. The vertical grey, orange, and purple ticks represent the Bragg positions for $\beta$-Cu$_2$V$_2$O$_7$, CuV$_2$O$_6$, and Cu$_{0.63}$V$_2$O$_5$, respectively.}
 \end{figure}

The powder sample of ZnCVO shows a pure phase with an identical structure with $\beta$-CVO, as shown by the Rietveld refinement on the x-ray diffraction patterns in Fig.~\ref{fig1} (a). This result is consistent with the previous work by Pommer {\sl et. al.,}~\cite{Pommer2003} where the Zn$_x$Cu$_{2-x}$V$_2$O$_7$ compound completely transformed to the $\beta$ phase at $x$~=~0.15. At lower doping concentrations, the samples show the mixed $\alpha - \beta$ phases, and the Zn concentration of $x$ =~0.15 is expectedly at the transition point. From the pure phase powder ZnCVO, the single-crystals of ZnCVO with the largest size of approximately 1 $\times$ 1 $\times$ 1 cm$^\text{3}$ ($m \approx$ 1.5 g) were obtained using the vertical gradient furnace. The natural cleaved facet can be identified as the crystallographic $a$-axis similar to the $\beta$-CVO single-crystals~\cite{He2008a}. The $\omega$-scan around $(020)$ Bragg peak using neutron scattering at BT7 (Fig.~\ref{order}, inset) with a Gaussian fit yields a full-width-at-half-maximum (FWHM) equal to 0.38(4)$^\circ$, indicative of good crystallinity. Rietveld refinements on the powder x-ray diffraction pattern obtained from the ground single-crystals, as shown in Fig.~\ref{fig1}~(b), can also be fitted well with the reported $\beta$-CVO crystal structure~\cite{Hughes1980}. In addition, powder neutron diffraction on ZnCVO and $\beta$-CVO powder samples were also performed at 30 K and 2.5 K for crystal structure and magnetic structure determination, respectively. At 30 K, the powder neutron diffraction patterns of both ZnCVO and $\beta$-CVO were refined against the reference $\beta$-CVO crystal structure as shown in Fig.~\ref{fig2}. Despite the presence of Zn, the diffraction pattern shows a pure $\beta$-CVO phase without any trace of other phases. The refined occupancy of the Cu site from powder neutron data in Table~\ref{tableBT1} yields 0.97(1) suggesting that the doping concentration of Zn is approximately 3\%, much lower than the stoichiometric ratio of 7.5\%. The powder neutron diffraction pattern of $\beta$-CVO, on the other hand, shows some impurity peaks which can be indexed with CuV$_2$O$_6$~\cite{CuV2O6} ($\approx$~9\%) and Cu$_{0.63}$V$_2$O$_5$~\cite{Cu063V2O5} ($\approx$ 6\%). The refined parameters obtained from both x-ray and neutron diffractions are summarized in Table~\ref{tableXRD} and  \ref{tableBT1}, respectively.

\begin{table}
\caption{\label{tableXRD} Fractional coordinates of powder ZnCVO, powder $\beta$-Cu$_2$V$_2$O$_7$, and ground single-crystals of ZnCVO samples obtained from the Rietveld refinements on the x-ray diffraction patterns measured at room temperature as those shown in Fig.~\ref{fig1}.}
\centering
\begin{tabular*}{0.48\textwidth}{@{\extracolsep{\fill}}ccccc}
\hline
Atom & Site & $x/a$ & $y/a$ & $z/a$ \\
\hline
\multicolumn{5}{c}{Powder ZnCVO}\\
Cu  & 8f &0.3114(6)  & 0.0758(6)  &  0.5134(5)  \\
V & 8f &0.2283(7) & -0.2261(6)  & 0.2889(6) \\
O(1) & 4e &0.0000 &  0.147(2)  & 0.7500  \\
O(2)  & 8f &0.265(2)  & -0.092(3)  & 0.621(2)  \\
O(3)  & 8f &0.364(2) &  -0.081(2)  & 0.383(2)  \\
O(4)  & 8f &0.247(2)  & 0.752(2)  & 0.869(1)  \\
\multicolumn{5}{c}{$a$ = 7.6802(2) \AA, $b$ = 8.0550(3) \AA, $c$ = 10.1118(3)}\\
\multicolumn{5}{c}{$\beta$ = 110.343(3)$^\circ$, $R_{p}$ = 6.02\%, $R_{wp}$ = 8.88\%}\\
\hline
\multicolumn{5}{c}{Powder $\beta$-Cu$_2$V$_2$O$_7$}\\
Cu  & 8f &0.3085(5)  & 0.0722(4)  &  0.5128(4)  \\
V & 8f &0.2264(8) & -0.2263(6)  & 0.2853(5) \\
O(1) & 4e &0.0000 &  0.151  & 0.7500  \\
O(2)  & 8f &0.264(2)  & -0.091(2)  & 0.636(2)  \\
O(3)  & 8f &0.373(2)  & -0.095(2)  & 0.396(2)  \\
O(4)  & 8f &0.238(2) &  0.753(2)  & 0.873(1)  \\
\multicolumn{5}{c}{$a$ = 7.6950(6) \AA, $b$ = 8.0239(6) \AA, $c$ = 10.1056(6)}\\
\multicolumn{5}{c}{$\beta$ = 110.266(4)$^\circ$, $R_{p}$ = 5.36\%, $R_{wp}$ = 7.08\%}\\
\hline
\multicolumn{5}{c}{Ground single-crystals of ZnCVO}\\
Cu  & 8f &0.3101(3)  & 0.0740(2)  &  0.5139(2)  \\
V & 8f &0.2229(4) & -0.2236(3)  & 0.2876(3) \\
O(1) & 4e &0.0000 &  0.130(1)  & 0.7500  \\
O(2)  & 8f &0.266(1)  & -0.098(1)  & 0.634(9)  \\
O(3)  & 8f &0.383(1)  & -0.094(1)  & 0.397(9)  \\
O(4)  & 8f &0.227(1) &  0.751(1)  & 0.867(8)  \\
\multicolumn{5}{c}{$a$ = 7.6757(1) \AA, $b$ = 8.0586(2) \AA, $c$ = 10.1100(2)}\\
\multicolumn{5}{c}{$\beta$ = 110.368(2)$^\circ$, $R_{p}$ = 4.91\%, $R_{wp}$ = 6.38\%}\\
\hline
\end{tabular*}
\end{table}

\begin{table}
\caption{\label{tableBT1} Fractional coordinates of ZnCVO and $\beta$-Cu$_2$V$_2$O$_7$ powder samples obtained from the Rietveld refinements on the powder neutron diffraction patterns measured at 30 K. Note that the large errors at the refined positions of vanadium are due to its weak neutron scattering cross section~\cite{Sears1992}.}
\centering
\begin{tabular*}{0.48\textwidth}{@{\extracolsep{\fill}}ccccc}
\hline
Atom & Site & $x/a$ & $y/a$ & $z/a$ \\
\hline
\multicolumn{5}{c}{ZnCVO}\\
Cu/Zn\footnote{The refined occupancy number for Cu atom is 0.97(1) and thus for Zn atom is approximately 0.03.} & 8f & 0.3123(2) &  0.0723(2) & 0.5149(2)   \\
V & 8f & 0.217(5) & -0.246(5)  & 0.293(3)  \\
O(1) & 4e & 0.0000 & 0.1327(3)  &  0.7500  \\
O(2)  & 8f & 0.2739(3) & -0.0951(3)  & 0.6345(2)   \\
O(3)  & 8f & 0.3802(3) & -0.0914(3)  &  0.3975(2)  \\
O(4)  & 8f & 0.2424(3) & 0.7536(3)  &  0.8738(2)  \\
\multicolumn{5}{c}{$a$ = 7.7131(1) \AA, $b$ = 8.0242(1) \AA, $c$ = 10.1292(2)}\\
\multicolumn{5}{c}{$\beta$ = 110.408(1)$^\circ$, $R_{p}$ = 4.27\%, $R_{wp}$ = 6.12\%}\\
\hline
\multicolumn{5}{c}{$\beta$-Cu$_2$V$_2$O$_7$}\\
Cu  & 8f & 0.3121(4) &  0.0698(4) & 0.5139(3)   \\
V & 8f & 0.232(8) & -0.267(8)  & 0.272(6)  \\
O(1) & 4e & 0.0000 & 0.1259(6)  &  0.7500  \\
O(2)  & 8f & 0.2811(5) & -0.0987(5)  & 0.6402(4)   \\
O(3)  & 8f & 0.3788(5) & -0.0877(5)  &  0.3991(3)  \\
O(4)  & 8f & 0.2456(6) & 0.7507(5)  &  0.8759(3)  \\
\multicolumn{5}{c}{$a$ = 7.7249(2) \AA, $b$ = 8.0013(2) \AA, $c$ = 10.1249(3)}\\
\multicolumn{5}{c}{$\beta$ = 110.315(2)$^\circ$, $R_{p}$ = 4.50\%, $R_{wp}$ = 6.09\%}\\
\hline
\end{tabular*}
\end{table}

\subsection{Magnetic structure}\label{sectionB}

Now we discuss the magnetic structure determinations on ZnCVO and $\beta$-CVO using powder neutron diffraction. As mentioned earlier in Section~\ref{sectionA}, we prepared both ZnCVO and $\beta$-CVO to confirm that they share not only crystal structure but also magnetic structure. We start with the irreducible representation analysis using the program \textsc{basirreps} in the \textsc{fullprof}~\cite{fullprof} suit. According to the crystallographic space group $C2/c$ with commensurate magnetic translation vector $\vec k = (0,0,0)$, there are four possible magnetic irreducible representations (IR) as described in Table~\ref{tableIRs}. The corresponding Shubnikov magnetic space groups for $\Gamma_1$, $\Gamma_2$, $\Gamma_3$, and $\Gamma_4$ are $C2/c$, $C2/c'$, $C2'/c'$, and $C2'/c$, respectively~\cite{shubnikov}.  With the assumption that ZnCVO and $\beta$-CVO have the same magnetic structure, we know that these systems undergo a paramagnetic to antiferromagnetic transition at the N{\'e}el temperature of $T_\text{N} \simeq$ 26 K which will be discussed in Section~\ref{sectionC}. When we start considering the exchange couplings along with the nearest-neighboring pairs $J_1$, there are two equivalent bonds between the Cu$^{2+}$ ions i.e., Cu1-Cu3 and Cu2-Cu4 (see Table~\ref{tableIRs} and Fig.~\ref{IRs}). It was originally believed that this system was the antiferromagnetic spin chain with alternating $J_1 - J_2$ bonds (not shown here). However, it has been later proposed using the DFT calculations~\cite{Tsirlin2010} that this system could be better described by the complex anisotropic honeycomb network. In their proposed model, the leading antiferromagnetic exchange interactions are along two $J_5$ and one $J_6$ (Fig.~\ref{IRs}) i.e., three bonds per site. However, there are still weak but non-negligible antiferromagnetic $J_1$ as well as the interplane $J_{14}$ couplings, making the spin network more complex than the simple honeycomb structure. It is therefore presumed that the Cu$^{2+}$ atoms must align antiparallel with their neighbors through the most prominent exchange interactions, here $J_1$, $J_5$, and $J_6$. In addition, the previous magnetization measurements by He {\sl et. al.,}~\cite{He2008b} on the single-crystals of $\beta$-CVO strongly suggested that the magnetic easy axis of this system was along the crystallographic $c$-axis. This suggests that the magnetic moment $m_a$ and $m_b$, despite their possible nonzero values, could be discarded.

\begin{table}
	\caption{\label{tableIRs}Magnetic irreducible representations (IR) and their basis vectors (BV) for Cu1($x,y,z$), Cu2($-x+1/2,-y+1,-z+3/2$), Cu3($-x+1/2,-y+1/2,-z+1$), and Cu4($x,-y+1,z-1/2$) (see Fig.~\ref{IRs}).}
	\centering
	\begin{tabular}{c c c c c c c c c c c c c c}
		\hline \hline
		&  & \multicolumn{3}{c}{Cu1} & \multicolumn{3}{c}{Cu2} & \multicolumn{3}{c}{Cu3}& \multicolumn{3}{c}{Cu4}\\
		\hline
		IR  & BV & $m_{a}$ & $m_{b}$ & $m_{c}$ &$m_{a}$ & $m_{b}$ & $m_{c}$ &$m_{a}$ & $m_{b}$ & $m_{c}$ &$m_{a}$ & $m_{b}$ & $m_{c}$\\
		$\Gamma_{1}$  & $\psi_{1}$ & 1 & 0 & 0 &    -1 & 0 & 0 &    1 & 0 & 0 &    -1 & 0 & 0\\
		 & $\psi_{2}$ & 0 & 1 & 0 &    0 & 1 & 0 &     0 & 1 & 0 &    0 & 1 & 0\\
		 &$\psi_{3}$ & 0 & 0 & 1 &    0 & 0 & -1 &     0 & 0 & 1 &    0 & 0 & -1\\
		
		$\Gamma_{2}$ & $\psi_{1}$ & 1 & 0 & 0 &    -1 & 0 & 0 &    -1 & 0 & 0 &    1 & 0 & 0\\
		 &$\psi_{2}$ & 0 & 1 & 0 &    0 & 1 & 0 &     0 & -1 & 0 &    0 & -1 & 0\\
		 &$\psi_{3}$ & 0 & 0 & 1 &    0 & 0 & -1 &     0 & 0 & -1 &    0 & 0 & 1\\
		
		$\Gamma_{3}$ & $\psi_{1}$ & 1 & 0 & 0 &    1 & 0 & 0 &    1 & 0 & 0 &    1 & 0 & 0\\
		 &$\psi_{2}$ & 0 & 1 & 0 &    0 & -1 & 0 &     0 & 1 & 0 &    0 & -1 & 0\\
		 &$\psi_{3}$ & 0 & 0 & 1 &    0 & 0 & 1 &     0 & 0 & 1 &    0 & 0 & 1\\
		
		$\Gamma_{4}$ & $\psi_{1}$ & 1 & 0 & 0 &    1 & 0 & 0 &    -1 & 0 & 0 &    -1 & 0 & 0\\
		&$\psi_{2}$ & 0 & 1 & 0 &    0 & -1 & 0 &     0 & -1 & 0 &    0 & 1 & 0\\
		&$\psi_{3}$ & 0 & 0 & 1 &    0 & 0 & 1 &     0 & 0 & -1 &    0 & 0 & -1\\
		
		\hline \hline
	\end{tabular}
\end{table}

\begin{figure}
\includegraphics[width=0.48\textwidth]{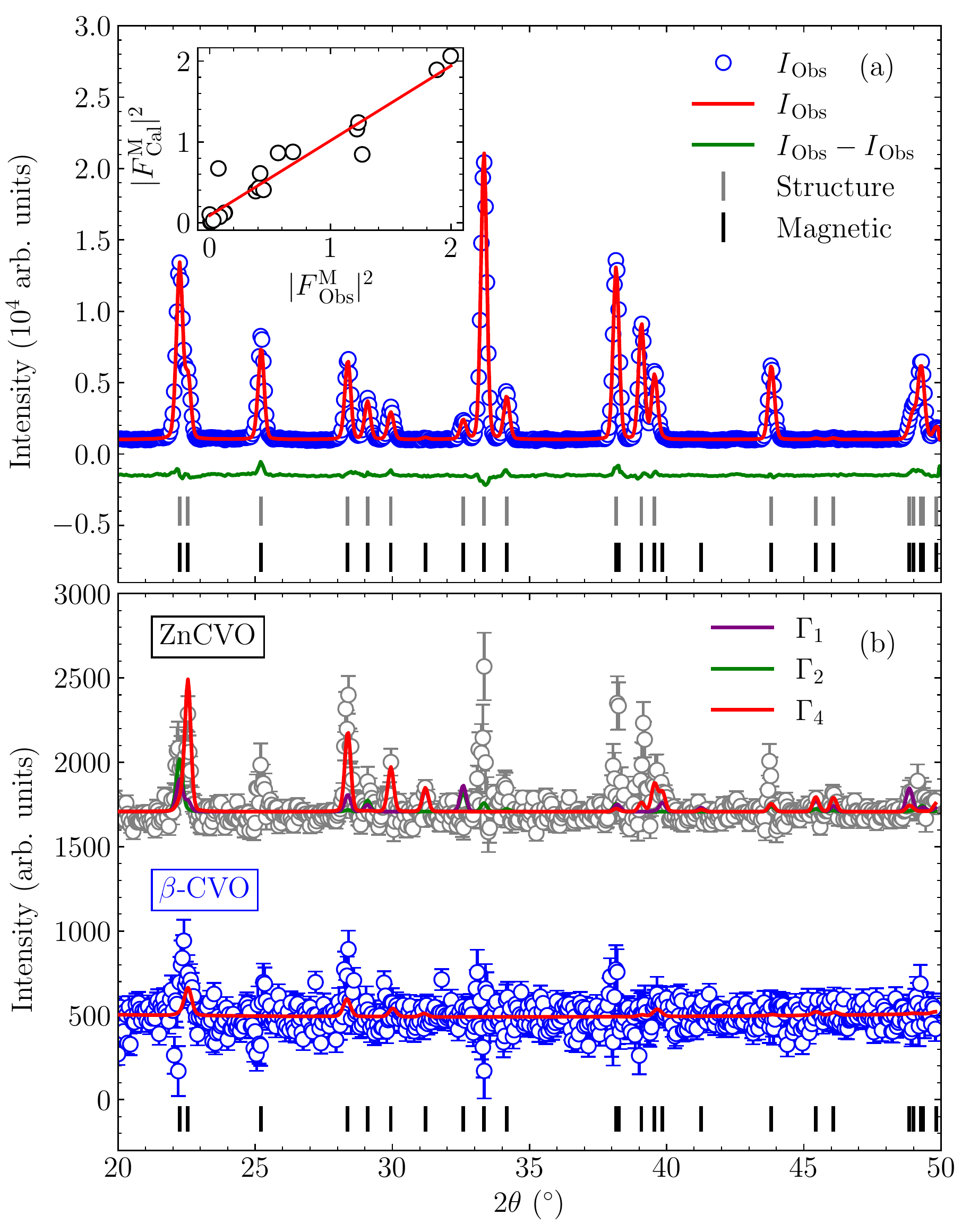}
 \caption{\label{fig4} (a) Powder neutron diffraction pattern with the Rietveld refinement to the magnetic structure $\Gamma_4$. Grey and black vertical marks represent the possible structure and magnetic Bragg positions, respectively. The inset shows the agreement between measured and calculated magnetic scattering intensities. (b) Powder neutron diffraction patterns at 2.5 K were subtracted by the 30 K data of ZnCVO (in grey circles) and $\beta$-CVO (in blue circles). The red lines are the subtraction between the Rietveld fits of the crystal structure at 30~K out of that magnetic structure at 2.5 K with $\Gamma_4$. The purple and green lines represent the same structural pattern subtracted from the magnetic pattern on ZnCVO using $\Gamma_1$, and $\Gamma_2$, respectively. The vertical black symbols represent the possible magnetic Bragg positions. The error bars represent three standard deviations throughout the article.}
 \end{figure}
 
Therefore among the four possible magnetic IRs, where we take into account the first nearest-neighbor couplings Cu1-Cu3 and Cu2-Cu4, we can rule out $\Gamma_1$ and $\Gamma_3$ where all spins align ferromagnetically along $m_a$. The reason that we pay attention to the first nearest neighbor is due to its strongest interaction as we will show later in Section~\ref{sectionD}. This leaves us with the two most probable magnetic IRs i.e., $\Gamma_2$ and $\Gamma_4$. It is obvious that only $\Gamma_4$ yields antiferromagnetic interaction on all neighboring bonds whereas $\Gamma_2$ gives ferromagnetic coupling on the fifth nearest neighbor. This assumption is based primarily on the DFT results by Tsirlin {\sl et. al.,}~\cite{Tsirlin2010} and Bhowal {\sl et. al.,}~\cite{Bhowal2017} (the citations will be omitted afterward when we mention the DFT results) where the predominant $J_1$, $J_5$, and $J_6$ bonds are all antiferromagnetic. With this initial analysis, we refined the powder neutron diffraction data at 2.5~K with $\Gamma_1$, $\Gamma_2$, and $\Gamma_4$ except for $\Gamma_3$ where the symmetry results in the ferromagnetic spin direction along $c$-axis. The magnetic structure of each IR is shown in Fig.~\ref{IRs}. The refined patterns from ZnCVO data with $\Gamma_1$, $\Gamma_2$, and $\Gamma_4$
are shown in Fig.~\ref{fig4} (b) for comparison along with their corresponding refined parameters summarized in Table~\ref{BT1}.

\begin{figure*}
\includegraphics[width = 0.9\textwidth]{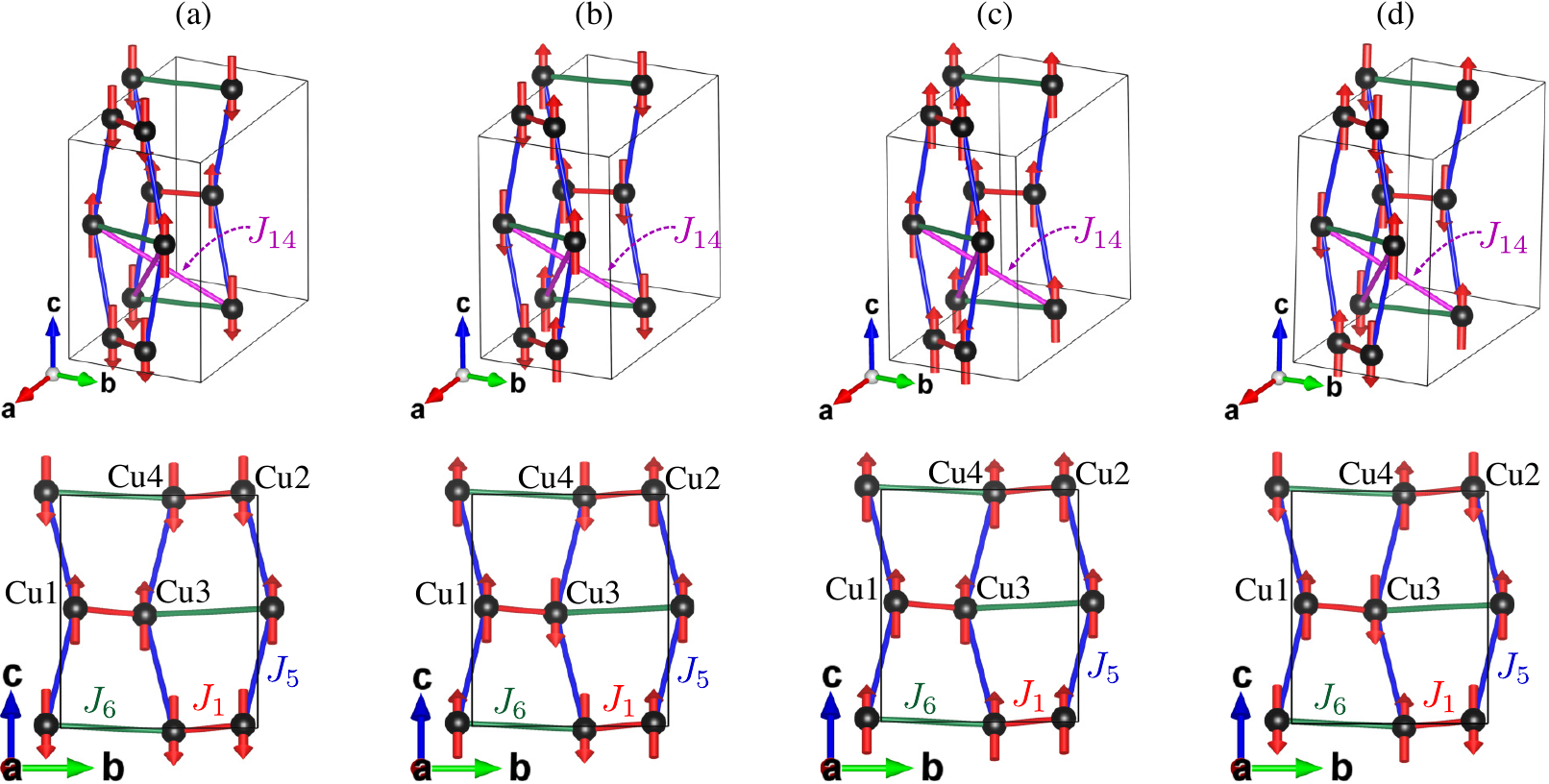}\hspace{7mm}

\caption{\label{IRs} Magnetic structure with irreducible representation (a) $\Gamma_1$, (b)  $\Gamma_2$, (c) $\Gamma_3$, and (d) $\Gamma_4$. When viewing along the crystallographic $b$-axis (upper row) the couplings $J_{14}$ (magenta) connect between the irregular honeycomb planes at which can be virtualized when view along the $a$-axis (lower row). The typical honeycomb structure is constructed from $J_1$ (red) and $J_5$ (blue) with the extra $J_6$ (green) bonds that connect between the opposite vertices.}
\end{figure*}

\begin{table}
 \caption{\label{BT1}The fitting parameters from the Rietveld refinements on the powder neutron diffraction of ZnCVO at 2.5 K.}
 \begin{tabular*}{0.47\textwidth}{@{\extracolsep{\fill}}cccc}
\hline
IRs &  $m_c$ ($\mu_B$) & $\chi^2$ & Magnetic $R$-factor\\ 
\hline  
$\Gamma_1$ & 0.6(2) & 11.6  &   26.5  \\
$\Gamma_2$ & 0.4(2) & 12.1 &   15.8  \\
$\Gamma_4$ & 0.72(9) & 9.1  &   13.1  \\ 
\hline
 \end{tabular*}
 \end{table}
 
It should be noted that the powder neutron diffraction patterns of both ZnCVO and $\beta$-CVO samples show very weak magnetic intensities, especially in $\beta$-CVO, and most of them are on top of the structural peaks. It is therefore very difficult to precisely extract the magnetic moment from the refinement. Furthermore, there could be large uncertainties in the refined values of the magnetic moment, and the exact magnetic structure could deviate from our proposed model. In order to present the magnetic intensities from the powder samples, we subtract the 30~K patterns from that of 2.5~K patterns, on both raw data and on the refined results, as shown in the low 2$\theta$ range in Fig~\ref{fig4} where the magnetic scattering is the most intense. It can be seen that the magnetic Bragg peak positions of both samples are consistent with the fitted model. Despite the dilution of the Cu sites by Zn, the magnetic intensities of ZnCVO are more pronounced than those of $\beta$-CVO where the intensities are most likely within the statistical error. Although we attempted to refine the magnetic structure on the $\beta$-CVO data we could not extract the magnetic moment with a reliable value. We could only obtain the magnetic moment from the ZnCVO data. The best fit is obtained from $\Gamma_4$ with the refined magnetic moment $m_c$ = 0.72(9)~$\mu_B$, the best among all three IRs. The refined pattern of ZnCVO with $\Gamma_4$ along with the plot of $\left| F^\text{M}_\text{Cal}\right|^2$ vs $\left| F^\text{M}_\text{Obs}\right|^2$ are shown in Fig.~\ref{fig4} (a). This magnetic structure will be further used in the spin-wave dispersion analysis in Section~\ref{sectionD}.

\subsection{Magnetic susceptibility}\label{sectionC}

Magnetic susceptibility of single-crystal ZnCVO was measured along two crystallographic axes i.e., $\chi_{\parallel a}$ with $H \parallel a$ (along the cleaved surface), and $\chi_{\perp a}$ with $H \perp a$ (parallel to the cleaved surface). The results, as shown in Fig.~\ref{chi}(a), reveal a broad peak at $T \approx$ 50 K indicating short-range correlations among the Cu$^{2+}$ spins. The paramagnetic upturn below $T \approx$ 20 K can be observed. This upturn, which corresponds to approximately 0.006$\mu_\text{B}$ at the field of 1 T and at the base temperature, is 13 times smaller than the ferromagnetism observed in $\alpha$-Cu$_2$V$_2$O$_7$~\cite{Gitgeatpong2017} and most likely a result of the presence of defective magnetic sites where Cu$^{2+}$ ions were substituted by Zn$^{2+}$ and thus the free Cu$^{2+}$ spins were produced~\cite{Pommer2003, Bag2021}. There is a large anisotropy between $\chi_{\parallel a}$ and $\chi_{{\perp a}}$ up to $T$~=~300 K similar to that observed in $\beta$-CVO by He {\sl et. al.,}~\cite{He2008b}. This unusual anisotropy was suggested as a result of the Jahn-Teller distortion~\cite{Jahn1937}. The similarity of the magnetic susceptibility behavior between ZnCVO in this work and $\beta$-CVO by the previous works, as well as our powder neutron diffraction data analysis, strongly suggest that both systems share the same magnetic properties.

\begin{figure}
\includegraphics[width = 0.47\textwidth]{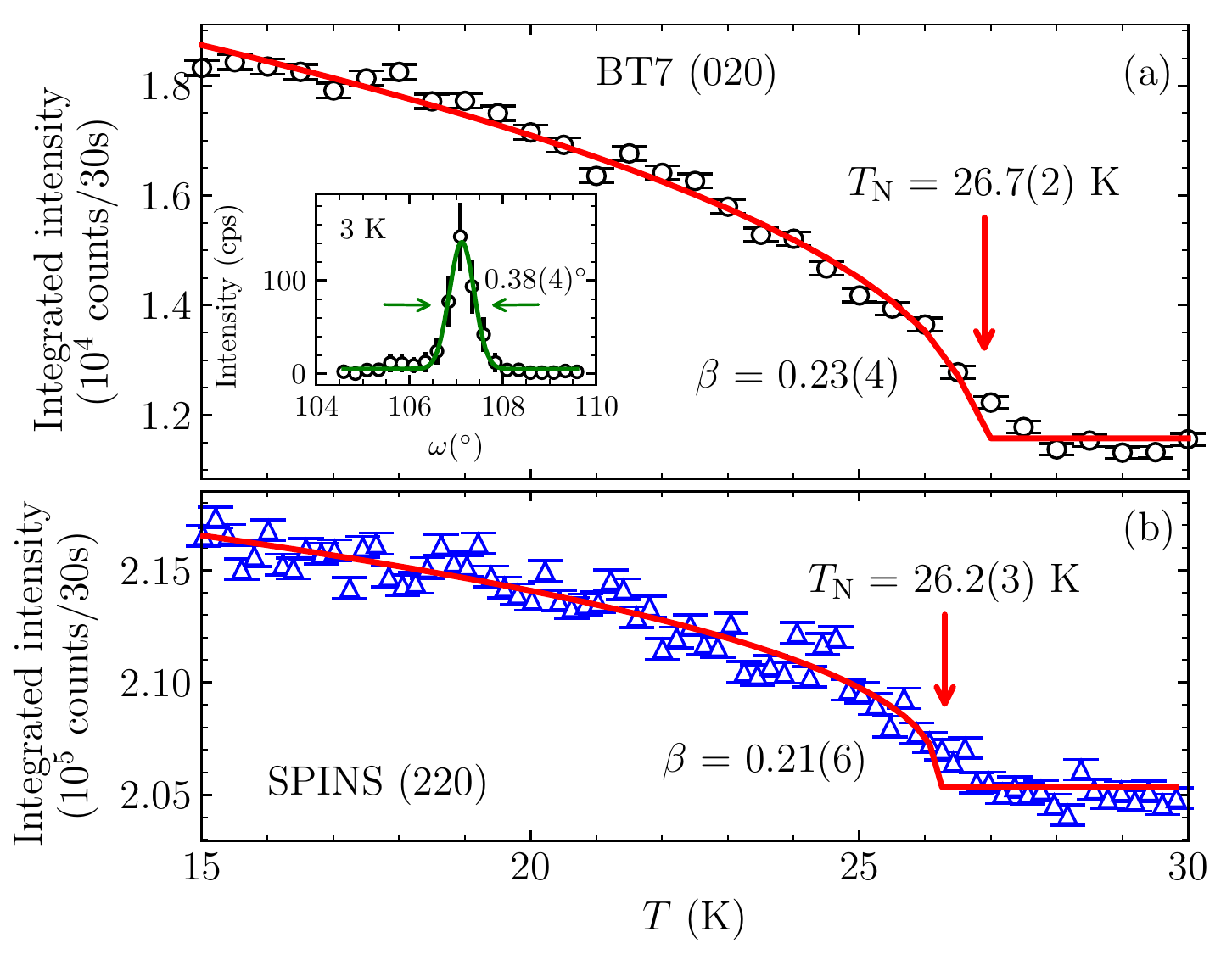}
\caption{Order parameter scans of the magnetic Bragg peaks (a) from BT7 spectrometer around (020) and (b) from SPINS spectrometer around (220). Red lines represent the critical exponent fits. Arrows indicate the N{\'e}el temperatures. Inset shows the omega scan around the (020) Bragg peak at $T$ = 3 K with the gaussian fit yielding FWHM = 0.38(4)$^\circ$.}\label{order}
\end{figure}

\begin{figure}
\includegraphics[width = 0.48\textwidth]{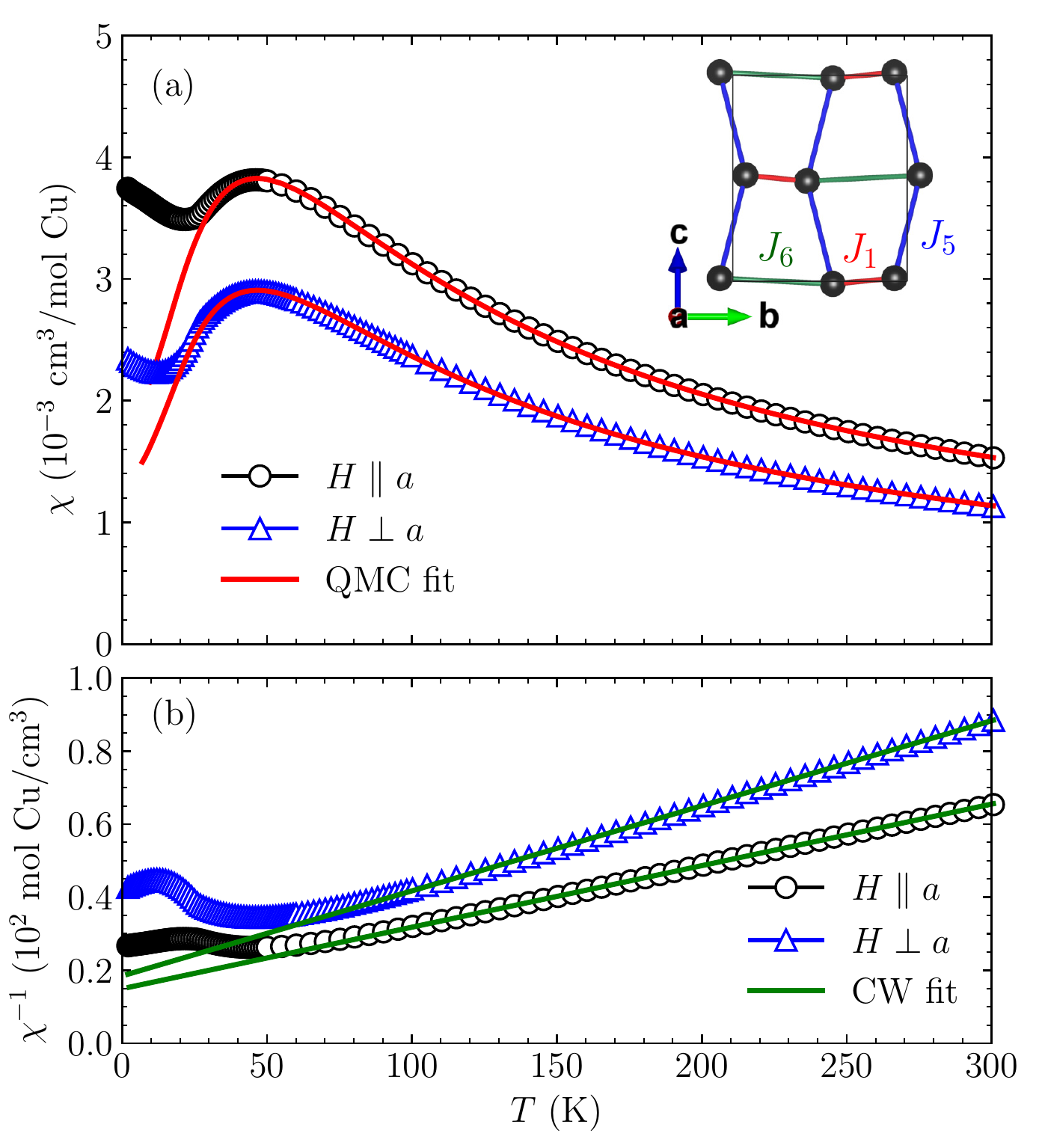}
\caption{(a) Magnetic susceptibility as a function of temperature with the magnetic field along crystallographic $a$-axis (blue triangle) and perpendicular to the $a$-axis (black circle). Red lines represent the QMC fits with the spin network as shown in the inset. (b) The inverse magnetic susceptibility and the Curie-Weiss law fit (green lines) at $T > 100$~K.}\label{chi}
\end{figure}

The plot of inverse magnetic susceptibility versus temperature, shown in Fig.~\ref{chi}(b), can be fitted well with the Curie-Weiss law ($\chi = C/(T-\Theta)$) at $T >$ 100 K. The fit yields the Curie-Weiss temperature of $\Theta$ = -79(1) K (-89(1) K) with $H\perp a$ ($H\parallel a$) indicating the dominant antiferromagnetic exchange interactions, and the Curie-Weiss constant $C$ = 0.429(1)~cm$^\text 3$K/molCu and 0.593(3)~cm$^\text 3$K/molCu for $H\perp a$ and $H\parallel a$, respectively. The effective magnetic moment can be estimated to $\mu_\text{eff} = \sqrt{3k_BC/N_A} = 1.852(4) \mu_B$ for $H\perp a$ and $2.17(1) \mu_B$ for $H\parallel a$. These values are slightly larger than the spin-only value of $\mu_\text{eff} = g\mu_B\sqrt{s(S+1)} = 1.73\mu_B$ for $g = 2$ and $S = 1/2$. The N{\'e}el temperature, $T_\text{N} \simeq$ 26~K is obtained from the exponent fit to the order parameter scans as a function of temperature on the magnetic Bragg peaks using elastic neutron scattering as shown in Fig.~\ref{order}. The fits were done in the range $15~\text{K} < T < 30~\text{K}$, close to the phase transition temperature, using equation $I = I_0\left( 1 - T/T_\text{N}\right)^{2\beta}$. This value is consistent with the observed $\lambda$-like transition at around 26 K from the heat capacity measurement on $\beta$-CVO single-crystal~\cite{He2008b}. The obtained critical exponent of $\beta \sim$ 0.2 is comparable to its cousin phase $\alpha$-Cu$_2$V$_2$O$_7$~\cite{Gitgeatpong2015}.

It should be noted that the doping of Zn on Cu sites results in a dilution of magnetic spin and typically decreases the N{\'e}el temperature~\cite{Eggert2002, Kataev2004, Bag2021}. However, in this case the value of $T_\text{N}$ is nearly the same as that of $\beta$-CVO~\cite{He2008a, He2008b}. In addition, the finite magnetic susceptibility below $T_\text{N}$ does not fit the paramagnetic impurity upturn as that observed in the powder sample by the previous works~\cite{Pommer2003}. The lattice parameters obtained from the Rietveld refinements (Table~\ref{tableXRD}) on the powder samples reveal that the lattice parameter $a$ slightly decreases while $b$, $c$, and the angle $\beta$ slightly increase upon the presence of Zn compared to pure $\beta$-CVO. Those lattice parameters on the ground single-crystals are also consistent with the powder ZnCVO sample. This suggests that the refined value of approximately 3\% Zn substitution on Cu sites only slightly alters the overall lattice parameters and does not affect the macroscopic magnetic properties.

To further estimate the average exchange interactions, we performed QMC simulations and fit the resulting simulated data to the broad peak of the magnetic susceptibility, provided that the spin network model and the values of leading exchange interactions were predicted. We proceed with the very first report on the DFT results (here we label the couplings according to the order of nearest-neighbor distances. The notation used by Tsirlin {\sl et. al.,} in Ref~\cite{Tsirlin2010} will be recalled in the parentheses). Among their various models, they suggested that the best realization of the spins network in $\beta$-CVO can be described by the fifth $J_5$ ($J_1$) and sixth $J_6$ ($J^\prime_1$) neighboring bonds, represented by the blue and green bonds, respectively, in Fig.~\ref{IRs} and the inset in Fig.~\ref{chi} (a). These two bonds connect the Cu$^{2+}$ ions into the irregular honeycomb network, i.e., three bonds per site, spanning the $bc$-plane when viewed along the $a$-axis. These honeycomb planes, according to the DFT results, are however not the perfect 2D since there are non-zero $J_{14}$ ($J_\perp$), represented by the magenta bonds in Fig.~\ref{IRs}, that connect between the adjacent honeycomb planes. There is also the suspicious $J_1$, formerly believed to be the leading exchange interaction, that appeared to be non-negligible from the DFT making the spin network in this system to be topologically the anisotropic magnetic 2D lattice (four bonds per site) with weak interplane couplings.

In our QMC simulation, we, therefore, construct the 2D spin network with anisotropic exchange interactions $J_1$, $J_5$, and $J_6$ as shown in the inset of Fig.~\ref{chi}~(a). The values of these couplings were obtained from the spin-wave dispersion fit on our inelastic neutron scattering data which will be discussed in Section~\ref{sectionD}. We simplify our spin network model by truncating the interplane fourteenth neighboring bond $J_{14}$ in the QMC simulation due to its very weak value. Although we fit the spin-wave dispersion based on the DFT model, the fitted parameters were obtained differently. Here we use the ratio $J_1 : J_5 : J_6$ = 1 : 0.61 : 0.25 for the QMC model. With $J_1-J_5-J_6$ interactions, the spin network resembles the irregular 2D edge-sharing trapezoid shape. We then conducted the QMC with the \textsc{loop} algorithm~\cite{loop} using the simulation package \textsc{alps}~\cite{alps}. The obtained QMC simulation result and the experimental magnetic susceptibility data were fitted using the equations,
\begin{equation}
\chi(T) = \chi_0 + \chi_\text{QMC}(T),
\end{equation}
with
\begin{equation}
\chi_\text{QMC}(T) = \frac{N_\text{A}\mu^2_Bg^2}{k_\text{B}J_\text{max}}\chi^*(t),
\end{equation}
where $N_\text{A}$, $\mu_\text{B}$, and $k_\text{B}$ are the Avogadro constant, Bohr magneton, and Boltzmann constant, respectively. The function $\chi^*(t)$ is the susceptibility as a function of reduced temperature $t = k_\text{B}T/J_\text{max}$ which was obtained by fitting the simulated QMC to the Pad{\'e} approximant~\cite{Johnston2000}. Here $J_\text{max}$ is $J_1$, the leading exchange interaction. The fitting parameters are the background $\chi_0$, the Land{\'e} $g$-factor, and the leading exchange interaction $J_\text{max}$ ($J_1$). The results are shown by the solid red lines in Fig.~\ref{chi}~(a) along with the two orthogonal magnetic field directions while the fitted parameters are summarized in Tabel~\ref{qmc}. The QMC simulation fits well with the magnetic susceptibility data over the broad maximum from $T \simeq$ 35 K up to 300~K yielding the leading exchange interaction $J_1 \simeq$~73~K ($\simeq$ 6.4 meV). Although the fitted values of the Land{\'e} $g$-factors are slightly deviated between $H \parallel a$ and $H \perp a$ data due most likely to the anisotropy, their average $g_\text{av}$ = 2.09(1) is still very close to the theoretical value of~2.

\begin{table}
 \caption{\label{qmc}The parameters obtained from the fit of QMC simulation to the magnetic susceptibility data when the field was applied along the crystallographic $a$-axis ($H \parallel a$) and perpendicular to the $a$-axis ($H \perp a$).}
 \begin{tabular*}{0.47\textwidth}{@{\extracolsep{\fill}}cccc}
\hline
Field direction &  $\chi_0~(\text{cm}^3/\text{mol Cu})$ & $J_1/k_B$ (K) & $g\text{-factor}$\\ 
\hline  
$H \parallel a$ & $1.9(1)\times10^{-4}$ &   $73.3(3)$ &   2.23(1) \\
$H \perp a$ & $1.0(1)\times10^{-4}$  &   $73.4(3)$  &   1.967(7)\\ 
\hline
 \end{tabular*}
 \end{table}

 \subsection{Spin-wave dispersion}\label{sectionD}

\begin{figure*}
\centering
\includegraphics[width = \textwidth]{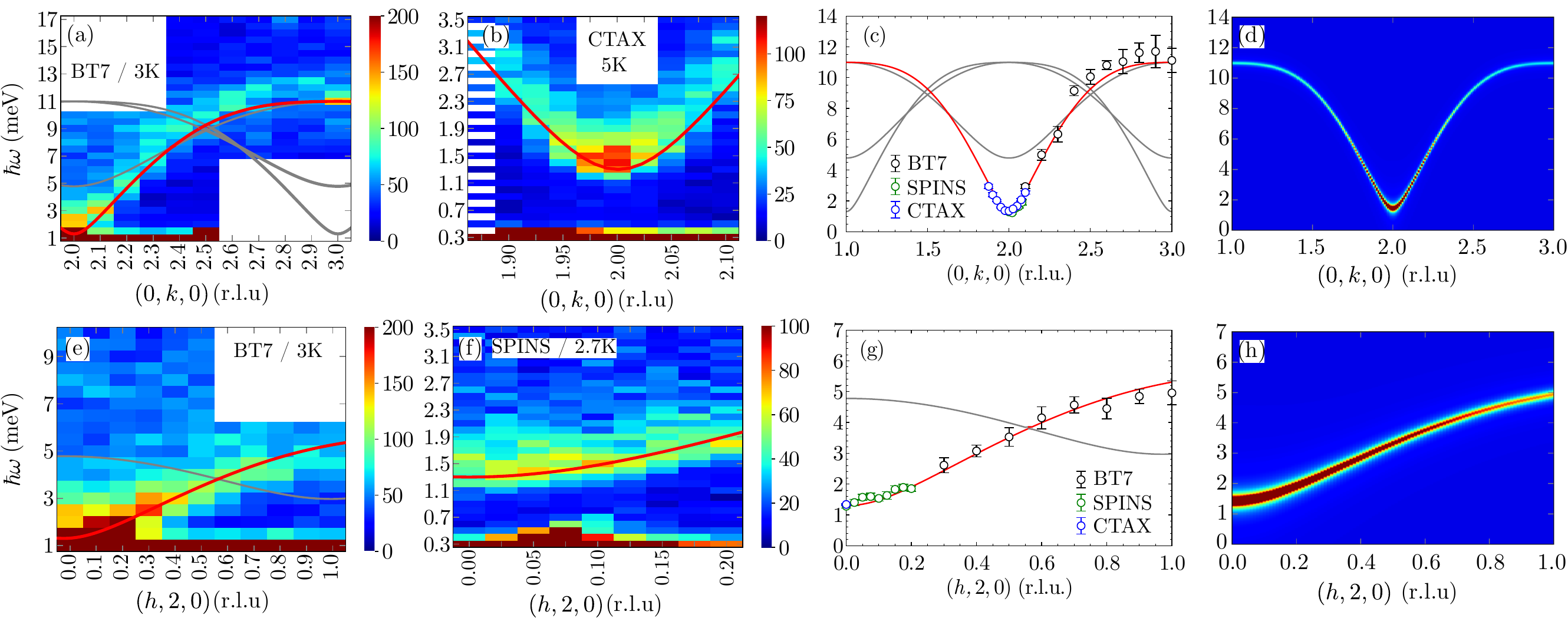}

\caption{Spin-wave dispersion of ZnCVO single-crystals along $(0,k,0)$ in (a) - (d) and along $(h,2,0)$ in (e) - (h). Red lines are the best fit for the dispersion relation. The intensity maps in (a), (b), (e), and (f) are plotted against the calculated curves. The fit between the model and the data is shown in (c) and (g). The calculated intensities are shown (d) and (h).}\label{spinwave}
\end{figure*}

All of the experimental data and analysis in the previous sections have led us to believe that the magnetic properties of ZnCVO could be a good realization of the $\beta$-CVO system. In this final section, we investigate the spin dynamics of ZnCVO single-crystals and analyze the obtained dispersion relation using the linear spin-wave theory (LSWT)~\cite{Anderson1952, Kubo1952}. We measured inelastic neutron scattering along two directions around the magnetic zone center at (0,2,0) i.e., along $(0,k,0)$ and along $(h,2,0)$. At low energy transfer ($\hbar\omega~<
~5$ meV), we conducted the experiments at SPINS and CTAX whereas at high energy transfer ($5$~meV $< \hbar\omega < 15$~meV) the experiments were done at BT7. The intensity map along the two directions at the base temperatures (depending on the spectrometer) is shown in Fig~\ref{spinwave} (a) - (b) and Fig~\ref{spinwave} (e) - (f). Figure \ref{spinwave} (a) shows the whole extent of the dispersion along $(0,k,0)$ from the magnetic zone center at (0,2,0) to the zone boundary at (0,3,0). The dispersion reaches its maximum at the energy transfer of $\approx$ 11 meV. At the magnetic zone center, we can see an energy gap clearly when using CTAX and SPINS spectrometers in Fig.~\ref{spinwave} (b) and (f) respectively. The dispersion is however different from its cousin phase $\alpha$-CVO where we found the splitting of the dispersion into two branches away from the magnetic zone center~\cite{Gitgeatpong2017}. This splitting, as mentioned earlier, was due to the presence of the DM interaction. On the other hand, in the $\beta$-CVO system which in this case is the ZnCVO, the crystal is centrosymmetric and thus DM interaction is absent due to the symmetry of the underlying crystal structure. Therefore, as we expected, the magnon dispersion in ZnCVO shows only one symmetric branch without the bidirectional shift. This evidence is a great test that the nonreciprocal magnon vanishes in the centrosymmetric crystal in the Cu$_2$V$_2$O$_7$ system. Along the $(h,2,0)$, on the other hand, the dispersion gradually increases from (0,2,0) up to the magnetic zone boundary at (1,2,0). This suggests that the spin interactions along the reciprocal lattice $b^*$ i.e., within the anisotropic lattice plane, are stronger than those along $a^*$ between the planes, and that the interaction between the planes should be relatively weak compared to the in-plane interactions.

In order to quantitatively analyze the exchange coupling values, we extract the dispersion relation from the convolute fit to the energy scan at each $Q$. The obtained dispersions from all data sets were plotted altogether as shown by the circle symbol in Fig~\ref{spinwave} (c) and (g) for $(0,k,0)$, and $(h,2,0)$ directions, respectively. Since the spin structure has been analyzed in Section~\ref{sectionC}, we need to construct spin interactions network for the LSWT fit. Again we start with the predicted models by the DFT calculations. In their work~\cite{Tsirlin2010}, they performed various computational approaches and showed that the leading exchange interactions were $J_5$ and $J_6$ forming the anisotropic honeycomb network with weak but non-negligible $J_1$ and the interplane $J_{14}$. We started with this model by introducing $J_1$, $J_5$, $J_6$, and $J_{14}$ into our spin model. The Hamiltonian that we used in our spin-wave fit is shown in Eq.~\ref{hamiltonian} below.

\begin{widetext}
\begin{equation}\label{hamiltonian}
\mathcal{H} = \frac{1}{2} \sum_{ij} \left\{ J_{ij} (\mathbf S_i \cdot \mathbf S_j) +  G_{ij}[\sin\beta (S_{zi}S_{zj} - S_{xi}S_{xj} - S_{yi}S_{yj}) + \cos\beta (S_{xi}S_{xj} - S_{yi}S_{yj} - S_{zi}S_{zj})]\right\},
\end{equation}
\end{widetext}
where $J_{ij}$ is the exchange interaction between spins $S_i$ and $S_j$, $\beta$ is the angle between the $a$-axis and $c$-axis due to the monoclinic system, and $G_{ij} = GJ_{ij}$, defined to be proportional to the exchange couplings, is the anisotropic parameter which gives rise to the spin gap at the magnetic zone center. We applied $J_{ij}$ and $G_{ij}$ to the first, fifth, sixth, and fourteenth neighboring bonds then fit the spin wave along both $(0,k,0)$ and $(h,2,0)$ directions simultaneously using least-square fitting routine to the modeled Hamiltonian.

Our result is, although qualitatively consistent with the DFT in terms of the representative leading exchange couplings, still quantitively deviated from the proposed honeycomb model. We note that for the data collected at BT7, despite its broad range covering from the magnetic zone center to the zone boundary, the resolution is rather low. It is possible that the exact values of the exchange interactions could slightly deviate from our results. As a result, despite the proposed honeycomb model with $J_5$ and $J_6$ as the leading exchange interactions, we instead get the largest value of 8.5(6) meV on $J_1$ which is also much higher than that in $\alpha$-CVO~\cite{Gitgeatpong2017}. The fitted results are shown by the red lines in Fig.~\ref{spinwave} whereas the fitted parameters are summarized in Table~\ref{dispersion}. These couplings yield the average in-plane exchange interactions $(J_1 + 2J_5 + J_6)/4$~=~5.3(2) meV. It should be noted that we also failed to fit our data when the second neighbor $J_2$ ($J_a$ in Ref~\cite{Tsirlin2010}) was introduced, in agreement with the DFT that this bond is rather weak and hence the previously proposed spin-chain model for this system is unfeasible. Although $J_{14}$ is rather weak compared to $J_1$ and $J_5$, this bond is non-negligible. This evidence leads us to conclude that the spin network of ZnCVO should be better described by the anisotropic 2D lattice with weak interplane couplings. Lastly, the calculated intensities of the dispersion along both $(0,k,0)$ and $(h,2,0)$ directions using the parameters in Table~\ref{dispersion} as shown in Fig.~\ref{spinwave} (d) and (h) can well describe the measured intensity maps.

\begin{table}
 \caption{\label{dispersion}The parameters obtained from the fit to the spin wave dispersions.}
 \begin{tabular*}{0.47\textwidth}{@{\extracolsep{\fill}}ccccc}
\hline
$J_1$ (meV) & $J_5$ (meV)  & $J_6$ (meV)  & $J_{14}$ (meV) & $G$ (meV) \\ 
\hline  
 8.5(6) & 5.3(3) & 1.9(4) & 0.5(1) & 0.0044(3) \\
\hline
 \end{tabular*}
 \end{table}
 
\section{Conclusion}\label{summary}

Our thorough x-ray and neutron diffractions have proved that ZnCVO is isostructural with $\beta$-CVO with a slight deviation in the lattice parameters. The large-sized single-crystals of ZnCVO can also be successfully grown from ZnCVO powder using the vertical gradient furnace. The system undergoes a paramagnetic to antiferromagnetic phase transition at $T_\text N \simeq$ 26 K. Magnetic structure determination using powder neutron diffractions suggested that, among the four possible magnetic irreducible representations, the diffraction pattern of ZnCVO can be best described by $\Gamma_4$ where the Cu$^{2+}$ spins anti-aligned with their neighbors along the crystallographic $c$-axis with the refined magnetic moment of $m_c$ = 0.72(9)$\mu_B$.

Magnetic susceptibility data of ZnCVO show large anisotropy between $H \parallel a$ and $H \perp a$ similar to the previous work on $\beta$-CVO. This suggests that not only does ZnCVO has the same crystal structure as $\beta$-CVO but they also share the same magnetic properties. The Curie-Weiss fit to the inverse magnetic susceptibility yields Curie-Weiss temperatures of $\Theta \simeq$  -80 K to -90 K (depending on the magnetic field direction) indicating the dominant antiferromagnetic exchange interactions. Our QMC simulation based on the spin-wave results can well reproduce the broad maximum on the magnetic susceptibility data.

Our inelastic neutron scattering data along $(0,k,0)$ and $(h,2,0)$ reveal typical symmetric spin-wave dispersion around the magnetic zone center, proving that the change from non-centrosymmetric to centrosymmetric crystal results in the absence of DM interaction and, thus, the nonreciprocal magnons. From DFT prediction and our magnetic structure results, we were able to fit the spin-wave dispersions data with the modeled spin Hamiltonian. Although the result is qualitatively consistent with the proposed $J_1-J_5-J_6-J_{14}$ model with strong coupling within the $bc$-plane and a rather weak interaction along $a^*$, the fitted values quantitatively deviate from the DFT calculations. Despite the proposed $J_5-J_6$ with weak $J_1$ and $J_{14}$ interactions, we obtained dominant $J_1-J_5$ with non-negligible $J_6$ and weak $J_{14}$. As a result, the network in ZnCVO resembles the anisotropic 2D lattice rather than the honeycomb lattice. These 2D spin networks are coupled through the weak interplane interaction $J_{14}$ resulting in the 3D ordered ground state.

\begin{acknowledgements}
G.G. would like to thank P. Limsuwan for his useful discussions. This work (Grant No. RGNS 63-203) was supported by the Office of the Permanent Secretary, Ministry of Higher Education, Science, Research and Innovation  (OPS MHESI), Thailand Science Research and Innovation (TSRI), and Phranakhon Rajabhat University. Work at Mahidol University was supported by the Thailand Center of Excellence in Physics and the National Research Council of Thailand (Grant No. N41A640158). P. S. was supported by the RGJ-PhD scholarship (Grant No. PHD/0114/2557) from Thailand Research Fund. We acknowledge the support of the National Institute of Standards and Technology, U.S. Department of Commerce, in providing the neutron research facilities used in this work. The identification of any commercial product or trade name does not imply endorsement or recommendation by the National Institute of Standards and Technology. A portion of this research used resources at the High Flux Isotope Reactor, a DOE Office of Science user facility operated by the Oak Ridge National Laboratory.
\end{acknowledgements}

\appendix

\bibliography{reference}

\end{document}